# Exact solution of the magnetic breakdown problem in quasi-one-dimensional geometry

**D. Radić [a], A. Bjeliš [a] and D. Zanchi [b]**

[a] Department of Physics, Faculty of Science, University of Zagreb, POB 162, 10001 Zagreb, Croatia ; e-mail: bjelis@phy.hr
[b] Laboratoire de Physique Théorique et Hautes Energies, 2 place Jussieu, 75252 Paris Cédex 05, e-mail : drazen@lpthe.jussieu.fr

**Abstract.** We present exact solution of the problem of electronic wave functions of quasi one-dimensional band with an inter-band gap at the Fermi surface and in the presence of magnetic field. The details of the analyzed model are appropriate to the situation in the Bechgaard salt $(TMTSF)_2ClO_4$ with the dimerizing anion order in the transverse direction. Limiting the effects of dimerization to the standard dimerization gap only, one obtains the electronic spectrum represented through solutions of a generalized Hill system of equations with simply periodic coefficients. The resulting wave-functions are discussed. In particular, we present the solutions for the case when the electrons spend as much time in the "junctions" as on their quasi-classical orbits. On the other hand, the limit when the tunnelling approach is valid is identified and the results are confronted with the well-known Slutskin-Kadigrobov solution. Furthermore, taking into account also the presumably finite transverse dimerizing displacements of chains, one encounters the qualitatively more complex problem of a system of equations with two-periodic coefficients. Some qualitatively new properties of electronic spectrum and corresponding one-electron physical quantities in this case will be discussed in detail.

**Key words.** anion gap, magnetic breakdown

## 1. MOTIVATION

The field-induced spin density wave (FISDW) state and other physical properties of Bechgaard salts with *one* open Fermi surface are interpreted by the well-known concept of one-dimensionalization of open electron orbits in the plane perpendicular to the external magnetic field $B$ [1]. However, in slowly cooled samples of $(TMTSF)_2ClO_4$ with the orientational anion ordering the Brillouin zone is reduced due to a finite dimerization potential along the *b*-axis perpendicular to the chain direction, and the electron spectrum is split into *two* sub-bands. Then one encounters the problem of magnetic breakdown of quasi-classical electron orbits through the barrier separating two Fermi surfaces. In the standard treatment [2,3] of this problem one constructs new electron states by including only the local tunnelling through the barrier at the Brillouin zone edges (transverse wave number close to $\pi/2b$). This approach is justified provided that the magnetic breakdown parameter $\kappa \equiv \omega_c t/V^2$ is (much) larger than unity [4]. Here $V$ is the potential generated by the anion ordering, $t$ is transverse hopping integral, and $\omega_c \equiv v_F eBb/\hbar$ is cyclotron frequency.

For $\kappa >> 1$ one has strong magnetic breakdown, i. e. the tunnelling probability exp $(-1/\kappa)$ is large. Evidently this regime is possible only if $V$ is weak with respect to $t$. However recent experimental [5] and theoretic [6, 7] results suggest that $V$ might be of the order or larger than $t$. In this case one cannot use the standard method of Refs. [2-4]. We introduce in the present paper an alternative exact approach, generally applicable to the problem of magnetic breakdown.

## 2. EXACT APPROACH

We start from the basic one-electron Hamiltonian with the Peierls substituted magnetic field,

$$H = iv_F\rho_3\partial_x + \tau_3 2t\cos(pb - Gx) + 2t'\cos 2(pb - Gx) - V\tau_1, \qquad (1)$$

where Pauli matrices $\tau$ and $\rho$ operate in the space of *left - right* and *bond - anti-bond* indices respectively, $t'$ is the effective parameter of imperfect nesting, and $G = \omega_c/v_F$ is the magnetic wave-number. Our aim is to find coefficients $\alpha$, $\beta$, $\theta$ in the unitary transformation ($|\alpha|^2 + |\beta|^2 = 1$)

$$\psi_+ = \exp(i\theta)(\alpha\varphi_A + \beta\varphi_B)$$
$$\psi_- = \exp(i\theta)(-\beta\varphi_A^* + \alpha\varphi_B^*) \quad (2)$$

from the fermion fields $\psi_\pm$ (here spin indices are omitted for simplicity) to new fields $\varphi_{A,B}$ for which the Hamiltonian (1) is reduced to the one-dimensional form

$$H = iv_F\rho_3\partial_x. \quad (3)$$

Noting that the coefficients $\alpha$, $\beta$, $\theta$ *depend on x and p only through the combination* $z \equiv pb - Gx$, one gets $\theta(z) = 2(t/\omega_c)$ sin $(2z)$ (like in the standard model without anion order), while $\alpha$ and $\beta$ satisfy a generalized Hill problem

$$i\omega_c\alpha'(z) = -2t\cos(z)\alpha(z) - V\beta^*(z)$$
$$i\omega_c\beta'(z) = -2t\cos(z)\beta(z) + V\alpha^*(z). \quad (4)$$

The solutions of this system of equations can be conveniently expressed in the Floquet form, $\alpha(z) = A(z) \exp(-iz\delta)$, $\beta(z) = B(z) \exp(iz\delta)$, where $A(z)$ and $B(z)$ are periodic with the period $2\pi$.

## 3. DISCUSSION

Floquet parameter $\delta$ and functions $A(z)$ and $B(z)$, which can be straightforwardly calculated numerically from Eq. 4 for any choice of physical parameters, give, together with the function $\theta(z)$, a complete characterization of all eigen-states for the problem (1). In particular $\delta$ enters into the one-dimensional spectrum of the diagonalized Hamiltonian (3) as the splitting parameter due to the finiteness of anion potential $V$,

$$E_{f,N}(k) = v_F[f(k-k_F) + GN] \pm v_F G\delta. \quad (5)$$

Here $f=\pm$ is the left-right index, while $N$ is an integer. Generally $\delta$ enters into phase factors in various response functions, *e.g.* into those appearing in the spin density wave susceptibilities relevant for the FISDW phase diagram of (TMTSF)$_2$ClO$_4$ [8]. On the other hand functions $A(z)$ and $B(z)$ regulate relative weights of e. g. branches in $E_{f,N}(k)$.

In Fig.1a we plot the dependence of the Floquet parameter $\delta$ on $V/\omega_c$ and $t/\omega_c$ and compare it with the corresponding approximate result of the standard local tunnelling treatment of Refs. [2-4] shown in Fig.1b. Obviously the latter approach ceases to be reliable in the regime $V \geq t$. This figure also covers the analytically reachable regime of weak anion potential and strong magnetic field ($\omega_c/t \gg 1$, $V/t \ll 1$) [9], which is however not reliable for relaxed (TMTSF)$_2$ClO$_4$. Then the Floquet parameter is given by $\delta \approx (V/\omega_c) J_0(4t/\omega_c)$.

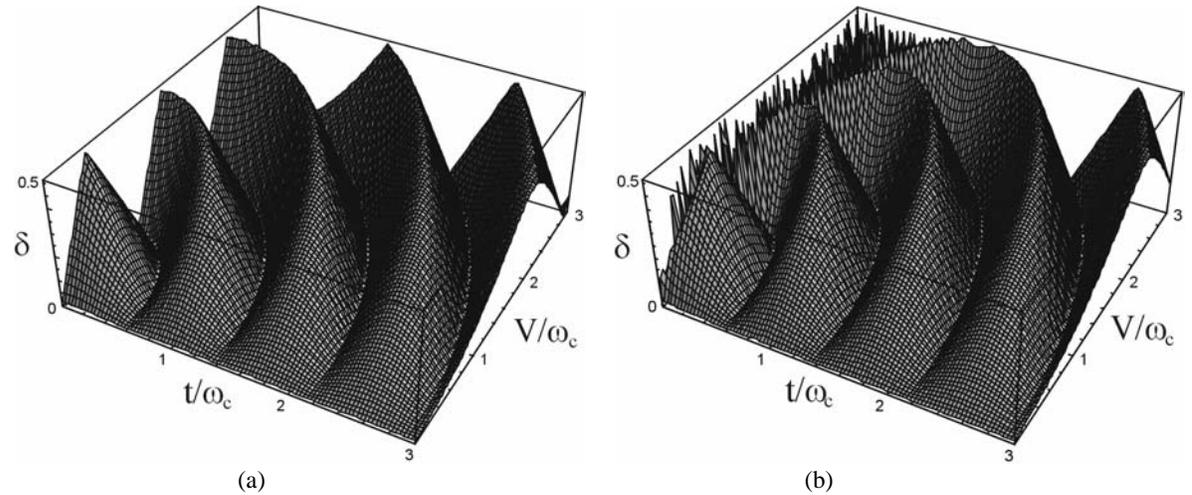

**Figure 1.** The present exact result (a) and the result of tunnelling approach [2-4] (b) for the dependence of Floquet parameter $\delta$ on $V/\omega_c$ and $t/\omega_c$.

In Fig. 2a we depict the dependence of the Floquet parameter $\delta$ on the inverse magnetic breakdown parameter $1/\kappa$ for few values of the ratio $V/t$, again compared with the corresponding result following from Refs. [2-4]. It is seen that the crossover from oscillating to saturating behaviour in $\delta$ coincides with the crossover from the weak ($\kappa < 1$) to the strong

($\kappa>1$) magnetic breakdown only for weak enough anion potential ($V < t$). In the opposite regime of strong anion potential ($V > t$) the position of the last zero of $\delta$ is not universal in $\kappa$. In fact, the universal oscillating behavior of the parameter $\delta$ can be represented in an alternative, more convincing way. Namely, as is seen in Fig.2b, the Floquet parameter $\delta$ shows universal oscillating behavior in the parameter $r \equiv [(\gamma V)^2 + t^2]^{1/2}/\omega_c$ with $\gamma \approx 0.77$ in the whole regime of physical parameters. The period of oscillations of $\delta$ in $r$ is approximately 0.8. We associate these oscillations with the rapid oscillations in $1/B$ with a frequency of 260 Tesla, observed in various transport and thermodynamic quantities in the metallic as well as in the FISDW state of relaxed (TMTSF)$_2$ClO$_4$ [10 - 12]. Choosing the characteristic values of parameters ($t = 300$K, $v_F = 2 \cdot 10^5$ m/s, $b = 7.7 \cdot 10^{-10}$ m) we get the best fit for $V \approx 0.8\, t$. This again indicates that the anion potential $V$ in the relaxed (TMTSF)$_2$ClO$_4$ is rather strong, in agreement with other independent evidences [5-7], but at variance with suggestions based on earlier X-ray measurements [13].

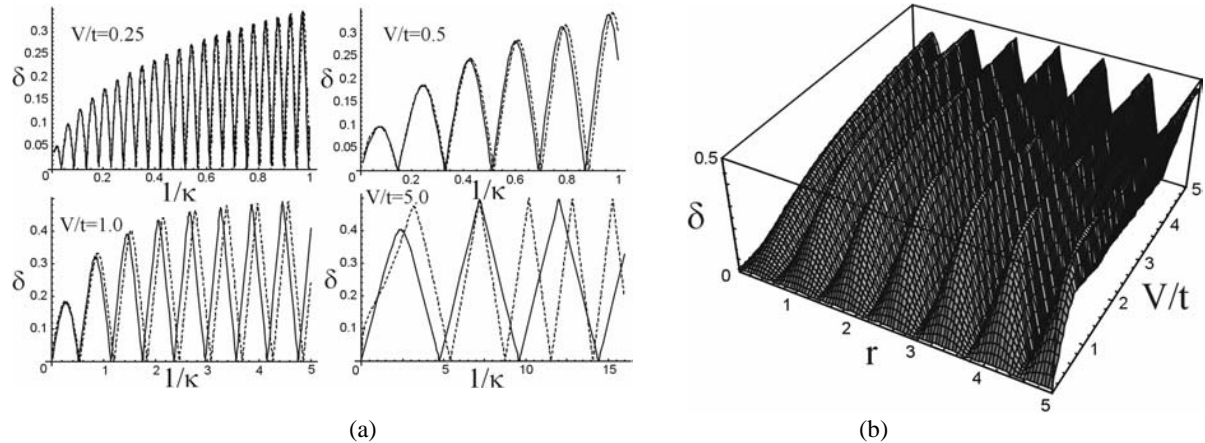

(a)          (b)

**Figure 2.** (a) Floquet parameter $\delta$ vs. inverse magnetic breakdown parameter $1/\kappa$ for $V/t$ = 0.25, 0.5, 1 and 5 for the present (full line) and the tunnelling (dashed line) approach. (b) Floquet parameter $\delta$ vs. parameter r and ratio $V/t$.

## 4. CONCLUSION

The present method of diagonalization of the Hamiltonian (1) as formulated by equations (2-4) is simultaneously an exact treatment of the quantum interference due to the magnetic field assisted tunnelling of quasi-classical orbitals through a barrier in the reciprocal space. It reproduces well the standard calculations of local tunnelling probabilities [2-4] in the regime of weak barriers in comparison with band corrugation ($V << t$ in our notation). The latter approach is not applicable in the intermediate regime ($V \approx t$) and in the regime of strong barriers ($V >> t$) in which our results collected in Fig. 2b suggest that the coherent tunnelling between two electron sub-bands is delocalized and takes place in the whole Brillouin zone.

The particular geometry of Fermi surfaces to which we apply the method (2-4) is that of (TMTSF)$_2$ClO$_4$ in relaxed state. As shown above the method enables a practically unambiguous fit of rapid oscillations observed in various physical properties in the metallic as well as in FISDW state. It leads to the additional evidence in favor of expectations that the anion potential $V$ is of the order of the transverse bandwidth $t$. This is in accordance with the analysis of SDW [6, 7] and FISDW [8] phase diagram for (TMTSF)$_2$ClO$_4$. Finally we note that the present method can be straightforwardly generalized to richer and more complex geometries and physical situations. One such example, a dimerization which induces band gap $V$ *and* finite relative displacements of two chain sub-families, will be discussed elsewhere.